\documentclass[pre,aps,twocolumn,superscriptaddress,nofootinbib]{revtex4-1}
\pdfoutput=1
\usepackage{amsmath, amsthm, amssymb}
\usepackage{amsfonts}
\usepackage{graphicx}
\usepackage{dcolumn}
\usepackage{bm}
\usepackage{textcomp}
\usepackage[normalem]{ulem}

\usepackage[titletoc,title]{appendix}

\usepackage{epsfig}
\usepackage{subfigure}
\usepackage{url}
\usepackage{float}
\usepackage{cases}

\usepackage{colordvi}
\usepackage[usenames,dvipsnames]{xcolor}

\usepackage[colorlinks=true, urlcolor=blue, anchorcolor=blue, citecolor=blue,filecolor=blue,linkcolor=blue,menucolor=blue
]{hyperref}

\usepackage[titletoc,title]{appendix}

\begin{document}
\title{Negative large deviations of the front velocity of $N$-particle branching
Brownian motion}

\author{Baruch Meerson}
\email{meerson@mail.huji.ac.il}
\affiliation{Racah Institute of Physics, Hebrew University of
Jerusalem, Jerusalem 91904, Israel}

\author{Pavel V. Sasorov}
\email{pavel.sasorov@gmail.com}
\affiliation{ELI Beamlines Facility, Extreme Light Infrastructure ERIC, 252 41 Dolni Brezany,  Czech Republic}

\begin{abstract}
We study negative large deviations of the long-time empirical front velocity of the center of mass of the one-sided $N$-BBM ($N$-particle branching Brownian motion) system in one dimension. Employing the macroscopic fluctuation theory, we study the probability that $c$  is smaller than the limiting front velocity $c_0$, predicted by the deterministic theory, or even becomes negative. To this end  we determine the optimal path of the system, conditioned on the specified $c$. We show that for $c_0-c\ll c_0$ the properly defined rate function  $s(c)$, coincides, up to a non-universal numerical factor, with the universal rate functions for front models  belonging to the Fisher-Kolmogorov-Petrovsky-Piscounov universality class. For sufficiently large negative values of $c$, $s(c)$ approaches a simple bound, obtained under the assumption that the branching is completely suppressed during the whole time. Remarkably, for all $c\leq c_*$, where $c_*<0$ is a critical value that we find numerically, the rate function $s(c)$ is \emph{equal} to the simple bound. At the critical point $c=c_*$ the character of the optimal path changes, and the rate function exhibits a dynamical phase transition of second order.
\end{abstract}
\maketitle

\section{Introduction}

The velocity  of reaction-diffusion fronts fluctuates because of the shot noise of the elemental processes of reactions and diffusion, see Ref. \cite{Panja} for an extensive review.
The front velocity fluctuations are especially significant, and therefore interesting, when the front propagates into an unstable state. The most interesting subclass of such fronts are the pulled fronts \cite{pulled}, which is the focus of our attention here. A simple model of pulled front is provided by the one-sided $N$-particle branching Brownian motion ($N$-BBM) in one dimension \cite{BDMM,Maillard,DerridaShi2016,DeMasi2019,Groisman2019,Berestycki2024}.  Here $N\gg 1$ independent Brownian particles with the diffusion constant $1$ branch at rate $1$ \cite{units}. At each branching event the leftmost particle is removed, so that the total number of particles is conserved. (Equivalently, the leftmost particle jumps to the location of any of the remaining $N-1$ particles chosen at random at rate $N-1$.) The $N$-BBM model belongs to a class of models proposed by Brunet and Derrida \cite{BDMM} to study selection mechanisms in biological systems.  In the deterministic limit $N\to\infty$,  this system develops a pulled front. The asymptotic velocity of the front $c_0=2$ is characteristic of a broad class of systems which belong to the universality class
of the Fisher-Kolmogorov-Petrovsky-Piscounov (FKPP) equation \cite{FKPP}
\begin{equation}\label{FKPPeq}
\partial_t u = u -u^2+\partial_x^2 u \,.
\end{equation}

This paper continues the line of work on negative ($c<2$)  large deviations
of the long-time empirical velocity $c$ of fluctuating pulled fronts of the FKPP type \cite{MSF,MVS}. Here we do it for the $N$-BBM model which, as we observe here, presents some new and surprising phenomena. In the microscopic formulation the empirical front velocity can be formally defined by the displacement of the position of the center of mass of the particles during a long time $T\gg 1$.  Exploiting the additional large parameter $N\gg 1$,  we employ an extended version of the macroscopic fluctuation theory (MFT) \cite{MFT,othernames} which accounts for particle reactions.  Relying on the MFT  \cite{MSF,MVS,MSK}, we will argue that, at $c<2$ and not too close to $2$ (that is, beyond the region of \emph{typical} fluctuations of $c$, see below), the long-time probability distribution $\mathcal{P} (c<2;T,N)$ exhibits a large-deviation behavior with respect to both $N$ and $T$:
\begin{equation}\label{LDbehavior}
-\ln \mathcal{P}(c<2;T,N) \simeq T N s(c)\,.
\end{equation}
The rate function per particle $s(c)$ can be formally defined as
\begin{equation}\label{rfdefinition}
s(c) =  - \lim_{\substack{N\to\infty \\ T\to\infty}} \frac{\ln \mathcal{P}(c;T,N)}{NT}\,,
\end{equation}
and it will be our main focus here {\cite{PTclarger2}.

The negative large deviations are determined by the optimal (that is, the most likely) density history of the system, which dominates the probability distribution in question \cite{MSF,MVS}. We argue that, in a finite range of empirical velocities, $c_*<c<2$,  the optimal path is described by traveling front solutions (TFSs) of the MFT equations. We find these solutions, and determine the rate function per particle $s(c)$ analytically, for $c=0$ and for $2-c\ll 2$. In the latter case we obtain

\begin{equation}\label{sc2}
s(c) \simeq K\,e^{-\frac{\pi}{\sqrt{2-c}}}\,,\quad 2-c\ll 2\,,
\end{equation}
where $K$ is a numerical constant which we determine. The rate function asymptotic~(\ref{sc2}) differs from the similar rate function asymptotics of Refs. \cite{MSF,MVS} only by the non-universal  constant $K$.

Farther from $c=2$, up to a negative critical value $c_*=-3.38\dots$,  we compute $s(c)$ by solving the MFT equations for the TFSs numerically. We observe that at $c=c_*$, $s(c)$ reaches a simple lower bound, obtained by assuming that the branching is completely suppressed during the whole time $T$, while the particles move the distance $cT$.

At $c<c_*$ we uncover a more involved asymptotic solution, where the optimal conjugate momentum field, as well as a very small fraction of the particles, travel faster than the center of mass of the system. For this optimal solution, which is \emph{not} a TFS,  the rate function $s(c)$ turns out to be identically equal to the simple bound, mentioned above, for all $c<c_*$. Because of the change of the character of the optimal path at the critical point $c=c_*$ the rate function $s(c)$ exhibits a dynamical phase transition of second order.

Here is a plan of the remainder of this paper. We briefly review the deterministic limit $N\to \infty$ of this model in Sec.~\ref{deterministic}. Section~\ref{fluctuations} is devoted to the calculations of the rate function $c(s)$.
We start with presenting the MFT formulation of the problem. Then we solve the MFT problem analytically and numerically, separately in the subcritical and supercritical regimes, and analyze the resulting dynamical phase transition. We briefly discuss our main results in Sec.~\ref{discussion}.

\section{Deterministic limit}
\label{deterministic}

In the limit of $N\to \infty$ the $N$-BBM model can be described by deterministic theory. Let us denote by $u(x,t)$ the particle density normalized by $N$. For the one-sided model that we are dealing with here the properly rescaled deterministic equations have the parameter-free form  \cite{DeMasi2019}
\begin{eqnarray}
&&\partial_t u (x,t) = \partial_x^2 u(x,t)+u(x,t)\,,\quad x>X(t)\,, \label{equ}\\
&&u(x,t)= 0\,, \quad x\leq X(t)\,, \label{BCdet} \\
&&\int_{X(t)}^\infty u(x,t) \,dx= 1\,.\label{norm}
\end{eqnarray}
Further, $u(x,t)$ is continuous at $x=X(t)$, and an initial condition must be specified. The effective ``absorbing wall" at $x=X(t)$ moves so as to keep the number  of particles $N$ constant at all times.

This deterministic system belongs to the FKPP universality class \cite{BDMM,Maillard,DeMasi2019,Groisman2019,Berestycki2024}. That is, at long times, $u(x,t)$ approaches the special limiting TFS $u(x,t) = U(x-2t)$ while $X(t) = 2 t$.
The function $U(\xi)$, where $\xi=x-2t$, obeys the ordinary differential equation (ODE)
\begin{equation}\label{ODENBBM}
U^{\prime\prime}+2U^{\prime}+U = 0,\quad \xi\geq 0\,,
\end{equation}
subject to the boundary condition $U(0)=0$ and the integral constraint $\int_{0}^\infty U(\xi) \,d\xi= 1$. The solution to this problem is unique:
\begin{equation}\label{U}
U(\xi) = \xi e^{-\xi}\,.
\end{equation}

\section{Velocity fluctuations and MFT}
\label{fluctuations}

\subsection{General}

The discreteness of particles and the stochastic character of elemental processes cause significant changes in the empirical front velocity of the pulled front of the FKPP type. These changes include a systematic negative shift of the mean front velocity  $\bar{c}$ from its deterministic value $2$ as well as fluctuations around the mean. The shift behaves, in the leading order,  as $2-\bar{c}\simeq \pi^2\ln^{-2} N$ \cite{BD1997}, whereas the variance of \emph{typical} fluctuations around $c=\bar{c}$ scales as $\ln^{-3} N$ \cite{Derridafluct2006}. Here we are interested in the left tail of the distribution
$\mathcal{P} (c;T,N)$ which describes  negative large deviations of $c$ \cite{pures}.

Before employing the MFT, let us obtain a simple lower bound on $\mathcal{P}(c;T,N)$ -- hence an upper bound on the rate function $s(c)$ -- by assuming that (a) there were no branching events altogether during the whole time $T\gg 1$, and (b) all of $N$ the Brownian particles moved beyond the required distance $cT$, corresponding to the empirical velocity $c$. The probability of event (a) -- a Poisson process, see \textit{e.g.} Ref. ~\cite{Gard}, with rate $N$ -- is equal to $e^{-NT}$. The
probability of event (b) follows from the asymptotic probability for a single Brownian particle to move a distance larger than $|c|T$ during time $T$. The latter probability is determined by the propagator of the diffusion equation, and at $T\gg 1$ it can be evaluated as $\exp(-c^2T/4)$. For $N$ independent particles this single-particle probability should be raised to the power $N$.  The product of the two probabilities yields the lower bound we are after:
\begin{equation}\label{probbound}
\mathcal{P}_{\text{b}}(c;T,N) \sim e^{-NT \left(\frac{c^2}{4}+1\right)}\,,
\end{equation}
which obeys the scaling behavior announced in Eq.~(\ref{LDbehavior}).
The corresponding upper bound on the rate function $s(c)$ is
\begin{equation}\label{sbound}
s_{\text{b}}(c) = \frac{c^2}{4}+1\,.
\end{equation}
As we will see,  for  $c<c_*$, where $c_*<0$ is introduced below, Eq.~(\ref{sbound}) yields the true rate function $s(c)$.

\subsection{MFT equations}

For the front to move slower than $c=2$ for a long time, a significant reconstruction of the particle density profile due to fluctuations is required. This reconstruction involves many particles. Therefore, the resulting probability density is expected to be exponentially small not only in time $T\gg 1$, but also in $N\gg 1$. The large parameter $N\gg 1$ justifies the application of the MFT \cite{MFT} extended to account for particle reactions \cite{Jona-Lasinio1,Jona-Lasinio2,Bodineau,EK,MS2011,MSF,MVS,MSK,Mmortal,MS2021,SVS2023}. For the $N$-BMM the MFT equations can be written in a Hamliltonian form \cite{MS2021}
\begin{eqnarray}
\partial_t q&=&q e^p + \partial_x \left(\partial_xq -2q\partial_x p\right)\,,
\label{q1}\\
\partial_t p&=& 1-e^p -\partial_x^2 p -\left(\partial_x p\right)^2 -\lambda(t)\,,
\label{p1}
\end{eqnarray}
for the optimal density field $q(x,t)$ and the ``conjugate momentum" density field $p(x,t)$. The corresponding constrained Hamiltonian is
\begin{equation}
\label{Hamiltonian}
H=H[q(x,t),p(x,t),\lambda(t)]= \int_{X(t)}^{\infty} dx\,\mathcal{H} (q,p,\lambda)\,,
\end{equation}
where
\begin{equation}\label{Ham}
\mathcal{H}(q,p,\lambda) = \mathcal{H}_0(q,p) +\lambda(t) q
\end{equation}
is the density of the constrained Hamiltonian, and
\begin{equation}
\label{H0am}
\mathcal{H}_0(q,p) =(e^p-1) q -\partial_x q\, \partial_x p
+q (\partial_x p)^2
\end{equation}
is the density of the unconstrained Hamiltonian. $\lambda(t)$ is a Lagrange multiplier, introduced to impose the particle conservation,
\begin{equation}\label{conservgeneral}
\int_{X(t)}^\infty q(x,t) \,dx= 1\,.
\end{equation}
see Ref. \cite{MS2021}.  This MFT problem is similar to the problems considered in Refs. \cite{MSF,MVS}. As we will see shortly, however, the conservation of the number of particles  and a simpler mechanism of limiting the particle proliferation introduce new elements, some of them surprising.  They also somewhat simplify the problem: for example, here Eq.~(\ref{p1}) is decoupled from the Eq.~(\ref{q1}).

It is useful to briefly discuss the physical meaning of the different terms in the Hamiltonian~(\ref{H0am}). The term $(e^p-1)q$ describes fluctuations of the branching rate: the branching rate is increased in comparison with its mean value $1$ at $p>0$ and decreased at $p<0$, see Eq.~(\ref{q1}).  The other two terms in the Hamiltonian~(\ref{H0am}) are the transport terms; they are familiar from the MFT of the gas of non-interacting random walkers (or, at large scales, of Brownian particles) \cite{MFT}. In particular,  the term $q (\nabla p)^2$  describes the fluctuational contribution to the particle flux, coming from the stochasticity of Brownian motion.

The boundary conditions for $q(x,t)$ and $p(x,t)$ at the absorbing wall are \cite{MS2021}
\begin{equation}\label{qat0}
q[x= X(t),t]=0\quad \text{and} \quad p[x=X(t),t]=0\,.
\end{equation}

The solution of the MFT problem for $q(x,t)$  describes the optimal path: the most likely density history conditioned on a certain $c$, whereas $p(x,t)$ describes the corresponding most likely realization of the noise.  In the absence of fluctuations one has $p=0$ and $\lambda(t)=0$. In this case Eq.~(\ref{p1}) holds trivially, and Eq.~(\ref{q1}) coincides with the deterministic equation~(\ref{equ}) which, at long times, describes a pulled traveling front with $c=c_0=2$.

To complete the MFT formulation, one should specify some initial and final conditions. However, when $T\gg 1$, details of these conditions are irrelevant in the leading order in $T$ \cite{MSF,MVS}. It suffices therefore to specify the distance $cT$ between the initial and final positions of the center of mass of the particles. Here $c$ is the empirical velocity.

Once the MFT problem is solved, the probability distribution $\mathcal{P}(c,N,T)$, can be evaluated, up to a preexponential factor, from the relation
\begin{equation}\label{probanonstat}
-\ln \mathcal{P}(c,N,T) \simeq N S(c,T)\,,
\end{equation}
where
\begin{equation}\label{actionmostgeneral}
S(c,T) = \int_0^T dt \int_{X(t)}^{\infty} dx\, \left(p \partial_t q -\mathcal{H}_0\right)
\end{equation}
is the Hamiltonian action per particle. Plugging Eqs.~(\ref{q1}) and (\ref{H0am}) into Eq.~(\ref{actionmostgeneral}), we obtain $S$ in terms of an integral along the optimal path:
\begin{equation}\label{actionnonstat}
\!\!S(c,T)=\!\!\int_0^T dt \int_{X(t)}^{\infty}dx \left[q \left(p e^p -e^p+1\right) + q (\partial_x p)^2 \right]\,.
\end{equation}
As $T\to \infty$, the action per particle $S(c,T)$ is expected to behave as $T s(c)$, where $s(c)$ is the rate function we are mostly interested in.

\subsection{Travelling front solution}

We will start with the assumption that, similarly to other models of the FKPP universality class \cite{MSF,MVS}, the leading-order result, for $T\gg 1$, is provided by a TFS of the MFT equations of the form
\begin{equation}\label{TFSsimple}
q(x,t) =q(x-ct)\quad \text{and} \quad p(x,t) = p(x-ct)
\end{equation}
with the empirical velocity $c$ (for the both fields $q$ and $p$) that we condition the process on. In this solution the Lagrange multiplier is time-independent, $\lambda(t) = \lambda =\text{const}$, and it plays the role of a ``nonlinear eigenvalue" of the problem which should be determined for a given $c$.

It is advantageous to go over to the Hopf-Cole variables $Q$ and $P$, obtained via the canonical transformation
$Q=q e^{-p}$ and $P=e^p-1$.  In the new variables the traveling front ansatz (\ref{TFSsimple}) in Eqs.~(\ref{q1}) and (\ref{p1}) yields two nonlinear ODEs:
\begin{eqnarray}
Q^{\prime\prime}+c Q^{\prime}
 +(2P+1+\lambda)Q&=&0\,, \label{1B}\\
 P^{\prime\prime}-c P^{\prime}
 +(P+1)(P+\lambda)&=&0\,, \label{2B}
\end{eqnarray}
where here and everywhere in the following $\xi =x-ct$.  The boundary conditions at $x=X(t)$ become
\begin{equation}\label{BC0}
Q(\xi=0) = P(\xi=0) = 0\,,
\end{equation}
and the mass conservation reads
\begin{equation}\label{conservQ}
\int_{0}^{\infty} Q(\xi) [P(\xi)+1]\, d\xi =1\,.
\end{equation}
The boundary condition for $P(\xi)$ at $\xi \to \infty$ is
\begin{equation}\label{BCPinfinity}
P(\xi\to \infty)=-1\,;
\end{equation}
it follows
from the fact that, in the original variables, $p(\xi \to \infty) = -\infty $ \cite{MSF}.

The asymptotic behavior of $Q(\xi)$ as $\xi\to \infty$ is determined by the demand that $q(\xi) = Q(\xi)[1+P(\xi)]$ goes to zero there.
Linearizing Eqs.~(\ref{1B}) and~(\ref{2B}) around $P=-1$ and demanding that $q(\xi\to \infty)=0$, we see that $1+P(\xi\to \infty)$ must decay as
\begin{equation}\label{Pdecay}
  1+P(\xi\to \infty) \sim \exp\left[\left(\frac{c}{2}-\sqrt{\frac{c^2}{4}+1-\lambda}\right)\xi\right]\,,
\end{equation}
whereas $Q(\xi\to \infty)$ behaves as
\begin{equation}\label{Qbehavior}
  Q(\xi\to \infty) \sim \exp\left[-\left(\frac{c}{2}+\sqrt{\frac{c^2}{4}+1-\lambda}\right)\xi\right]\,.
\end{equation}

When $c>0$, Eq.~(\ref{Qbehavior}) describes a decay. A decay also occurs for negative $c$ which are sufficiently small by the absolute value.  However, for sufficiently large negative $c$, $\lambda$ becomes larger than $1$, and the exponential decay of $Q(\xi)$ gives way to exponential growth. We found numerically that this change of behavior occurs at $c\simeq -1.079$ (see Fig. \ref{snum} below). This regime change, however, does not affect the optimal particle density $q(\xi)$, which still decays at infinity as it should.

The traveling front ansatz (\ref{TFSsimple}) implies that the action per particle (\ref{actionmostgeneral}) is indeed proportional to time: $S(c,T)=T s(c)$ where, in the new variables,
\begin{equation}\label{snewvar}
s(c) = \int_0^{\infty}  d\xi \left[P^{\prime}Q^{\prime}-Q(P+P^2)\right]
\end{equation}
is the rate function. Integrating by part the term $P^{\prime}Q^{\prime}$ in Eq.~(\ref{snewvar}) and using the boundary conditions (\ref{BC0}), we obtain
\begin{equation}\label{seqlambda}
s(c) = \lambda\,.
\end{equation}
That is, the calculation of the rate function in the travelling front regime requires  only  determining the Lagrange multiplier $\lambda$ as a function of $c$. Equation~(\ref{seqlambda}), along with Eqs.~(\ref{Pdecay}) and~(\ref{Qbehavior}), also imply that the TFSs can exist only for $0<\lambda<c^2/4+1$, as we indeed find here. This also means that when TFSs exist, they automatically obey the restriction $s(c)<s_{\text{b}}(c)$, where $s_{\text{b}}(c)$ is described by Eq.~(\ref{sbound}).

The traveling front problem is now fully defined. Importantly, Eqs.~(\ref{1B}) and (\ref{2B}) possess a conservation law:
\begin{equation}\label{ODEcons}
Q^{\prime} P^{\prime}+ Q(P+1)(P+\lambda)=\text{const}\,,
\end{equation}
where the constant on the right hand side vanishes by virtue of the asymtptotic behaviors of $P$ and $Q$ at infinity, see Eqs.~(\ref{Pdecay}) and (\ref{Qbehavior}).  (Similar conservation laws for TFSs in reacting and diffusing gases were encountered in Refs.~\cite{MSF,MSK,MS2021}.) Then, using Eq.~(\ref{BC0}), we obtain an important  -- and convenient -- additional  condition at $\xi=0$:
\begin{equation}\label{Pder}
P^{\prime}(\xi = 0) = 0\,.
\end{equation}

\subsubsection{$2-c\ll 2$ asymptotics}
\label{closeto2}

The region of $2-c\ll 2$ is universal for pulled fronts of the FKPP type \cite{MSF,MVS}. Here one can use the asymptotic method developed in Ref. \cite{MSF}. The method is based on the observation that, as $c$ approaches $2$ from below, the $Q$- and $P$-profiles become widely separated in space. This feature is evident in Fig. \ref{PQseparation} which shows the $Q$- and $P$-profiles for $c=1.97$, which we obtained numerically.  The asymptotic method \cite{MSF} in its general form involves obtaining approximate analytical solutions for $P$ and $Q$ in two regions: region (i) where $|P| \ll 1$, and region (ii) where $Q\ll 1$, and a subsequent matching of these solutions in the \emph{joint} region (iii)  where \emph{both} strong inequalities $|P| \ll 1$ and $Q\ll 1$ hold. In the present problem the calculations are somewhat shorter, because it suffices to solve only Eq.~(\ref{2B}) for $P(\xi)$ and, in the process of solving it, determine $\lambda=\lambda(c)$ and therefore the rate function $s(c)=\lambda$.

\begin{figure}
\includegraphics[width=0.4\textwidth,clip=]{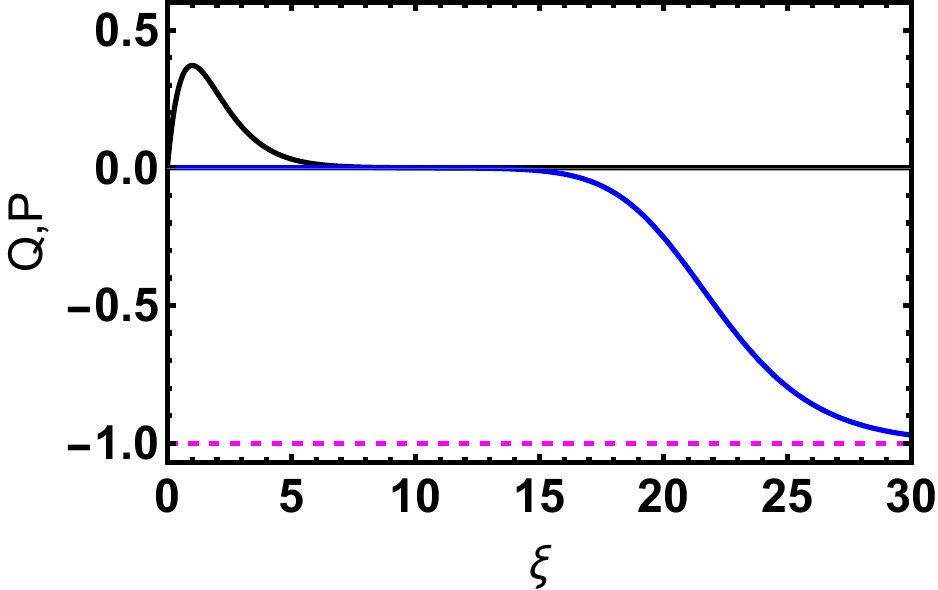}
\caption{$Q(\xi)$ (black) and $P(\xi)$ (blue), computed numerically for $c=1.97$. The spatial separation of the profiles is clearly seen.}
\label{PQseparation}
\end{figure}

In the region of $|P|\ll 1$ (the left region) we can linearize Eq.~(\ref{2B}) with respect to $P$:
\begin{equation}\label{2Blin}
P^{\prime\prime}-c P^{\prime}
 +P+\lambda=0\,.
\end{equation}
The solution of this linear equation, obeying the boundary conditions $P(0)=P^{\prime}(0)=0$, is
\begin{eqnarray}
 P^{\text{left}}(\xi)&=&-\frac{c \lambda  e^{\frac{c z}{2}} \sin
   \left(\sqrt{1-\frac{c^2}{4}}
   z\right)}{\sqrt{4-c^2}} \nonumber \\
  &+& \lambda  e^{\frac{c z}{2}} \cos
   \left(\sqrt{1-\frac{c^2}{4}} z\right)-\lambda\,. \label{Pleftsol1}
\end{eqnarray}
It can be simplified by using the small parameter $\delta\equiv 2-c \ll 2$:
\begin{equation}\label{Pleftsol2}
 P^{\text{left}}(\xi) \simeq -\frac{\lambda}{\sqrt{\delta}} e^{\xi} \sin (\sqrt{\delta} \xi)+\lambda e^{\xi}\cos (\sqrt{\delta} \xi)-\lambda\,.
\end{equation}

In the region of $|P|=O(1)$ (the right region) we can neglect in Eq.~(\ref{2B}) the term $\lambda$ which is expected to be very small, see the resulting Eq.~(\ref{sc2a}). The resulting equation,
\begin{equation}\label{Pright}
P^{\prime\prime}-c P^{\prime}+P+P^2=0\,,
\end{equation}
can be mapped into the ODE which describes the TFSs, $u(x,t)=u(x-ct)$, of the deterministic FKPP equation~(\ref{FKPPeq}):
\begin{equation}\label{FKPP1}
u^{\prime\prime}+c u^{\prime}+u-u^2=0\,.
\end{equation}
Indeed, let  $u(\xi)$ be the solution of Eq.~(\ref{FKPP1}) subject to the standard boundary conditions $u(-\infty)=1$
and $u(\infty) =0$. For a given $c$ this solution is unique up to translation in $\xi$. Then the function
\begin{equation}\label{mapping}
P^{\text{right}}(\xi) = - u(\xi_0 - \xi)
\end{equation}
with arbitrary $\xi_0$ solves our Eq.~(\ref{Pright}) with the boundary conditions $P(\infty)=-1$ and $P(-\infty)=0$.

\begin{figure}[t]
\includegraphics[width=0.4\textwidth,clip=]{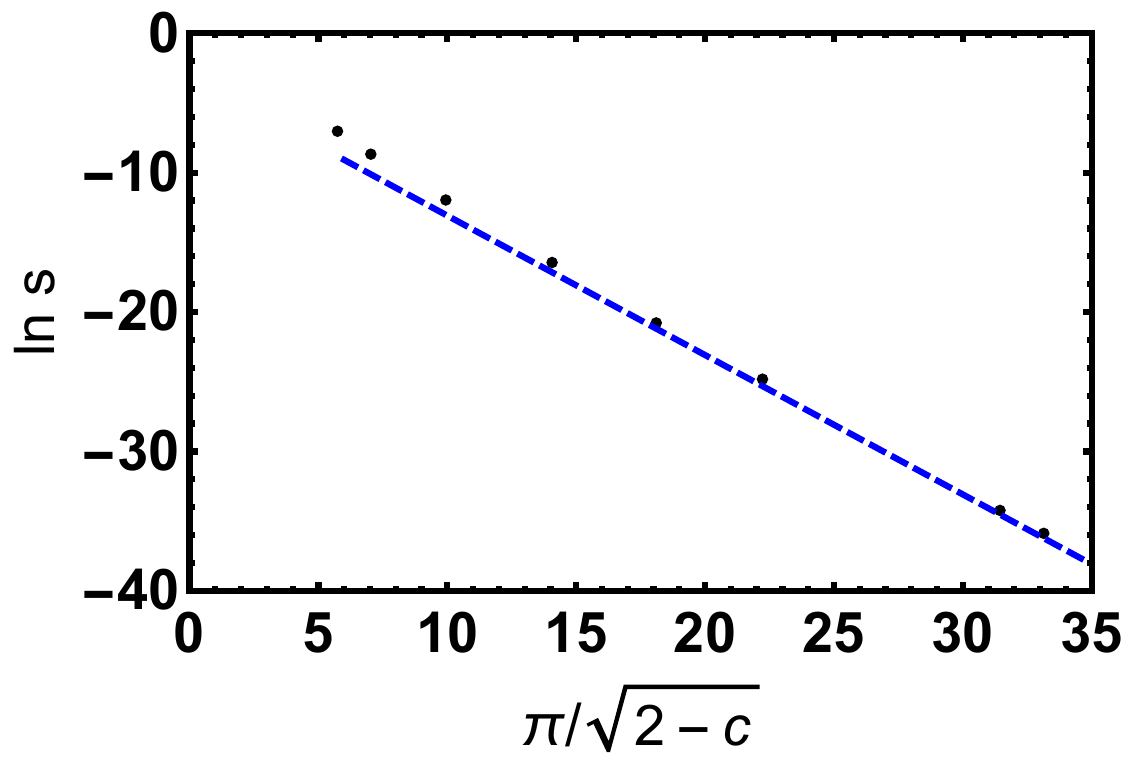}
\caption{$\ln s$ versus $\pi/\sqrt{2-c}$ for $c$ close to $2$. Black points: numerical results. Dashed
line: the asymptotic (\ref{sc2}).}
\label{sasympt}
\end{figure}

Crucially, the solution $P^{\text{left}}(\xi)$, described by Eq.~(\ref{Pleftsol2}),  can be matched,  in the joint region, with the ``left asymptotic"  of
$P^{\text{right}}(\xi)$. In its turn, the latter is simply related, through the mapping (\ref{mapping}),  to the large-argument asymptotic of $u(\xi)$. When $\delta \to 0$, this large-argument asymptotic has the form
\begin{equation}\label{uinfinity}
u(\xi\to \infty) \simeq \frac{B}{\sqrt{\delta}} e^{\xi} \sin(\sqrt{\delta} \xi)\,,
\end{equation}
where, in the leading order in $2-c\ll 2$, we can set the coefficient $B\simeq 0.142$ at its value in the asymptotic $u(\xi \to \infty) = B \xi e^{-\xi}$ of the TFS with $c=2$ \cite{aboutB}.   Performing the matching by using Eqs.~(\ref{Pleftsol2}), (\ref{mapping}) and (\ref{uinfinity}), we finally obtain $\xi_0\simeq \pi/\sqrt{\delta}+1 \gg 1$ and
\begin{equation}\label{sc2a}
\lambda(c)= s(c) \simeq \frac{B}{e}\,e^{-\frac{\pi}{\sqrt{2-c}}} \simeq 0.052\,e^{-\frac{\pi}{\sqrt{2-c}}}\,,\quad 2-c\ll 1\,,
\end{equation}
which coincides with Eq.~(\ref{sc2}) with $K\simeq 0.052$.  As one can see from Eq.~(\ref{sc2}), close to $c=2$ the rate function is very small. It differs from the similar rate functions of Refs. \cite{MSF,MVS} for $2-c\ll 2$ only by the non-universal  constant numerical factor $0.052$. Figure \ref{sasympt} compares, without any adjustable parameters, the prediction of Eq.~(\ref{sc2}) with our numerical results, to be described below, for $c$ close to $2$.

What are the applicability conditions of the asymptotic (\ref{sc2}) of the rate function? On the one hand, we have assumed that $2-c \ll 2$. On the other hand, the MFT is valid when there are many particles in the relevant front region at the leading edge of the front, where $\xi \simeq \xi_0 \simeq \pi/\sqrt{2-c}$. The latter condition leads to the same strong inequality $\bar{c} -c \gg 2 \pi^3 \ln^{-3} N$ as in Ref. \cite{MSF}. But again, for the formal definition of $s(c)$ in Eq.~(\ref{rfdefinition}) the logarithmic in $N$ region of $c$ shrinks to zero, and the applicability condition becomes simply $c<2$.

\subsubsection{Exact solution for $c=0$}
\label{exact}

$c=0$ is a large deviation where the optimal density $q(\xi)\equiv q(x)$ is stationary. This is the only case where we are aware of an exact solution of the traveling (or rather standing) front problem \cite{samesol}. The solution is the following, both in $Q,P$, and in $q,p$ variables:
\begin{eqnarray}
  Q(\xi) &=& \sqrt{\frac{8}{3}}
  \tanh \left(\frac{\xi}{\sqrt{6}}\right)\text{sech}^2\left(\frac{\xi}{\sqrt{6}}\right)\,, \label{Qsimple}\\
  P(\xi) &=&  -1+\text{sech}^2\left(\frac{\xi}{\sqrt{6}}\right)\,, \label{Psimple}\\
  q(\xi) &=& \sqrt{\frac{8}{3}}
  \tanh \left(\frac{\xi}{\sqrt{6}}\right)\text{sech}^4\left(\frac{\xi}{\sqrt{6}}\right)\,, \label{qsimple}\\
  p(\xi) &=& 2\,\ln \text{sech} \left(\frac{\xi}{\sqrt{6}}\right)\,. \label{psimple}
\end{eqnarray}
For this solution one obtains $\lambda=1/3$, so $s(c=0)=1/3$. Figure \ref{c=0_1} shows the optimal solutions for $Q$ and $P$ versus $\xi$ as described by Eqs.~(\ref{Qsimple}) and (\ref{Psimple}), respectively.  Figure \ref{c=0_2} compares the optimal density profile $q(x)$ with the deterministic TFS (\ref{U}). Because of a higher negative density gradient at small $\xi$ the particle loss at the boundary $\xi = 0$ is enhanced in comparison with the deterministic (that is, unconditioned) regime. This loss is balanced by the enhanced particle production via branching in the central region.  These two effects make it possible (but of course improbable!) to completely suppress the front motion.

\begin{figure}
\includegraphics[width=0.4\textwidth,clip=]{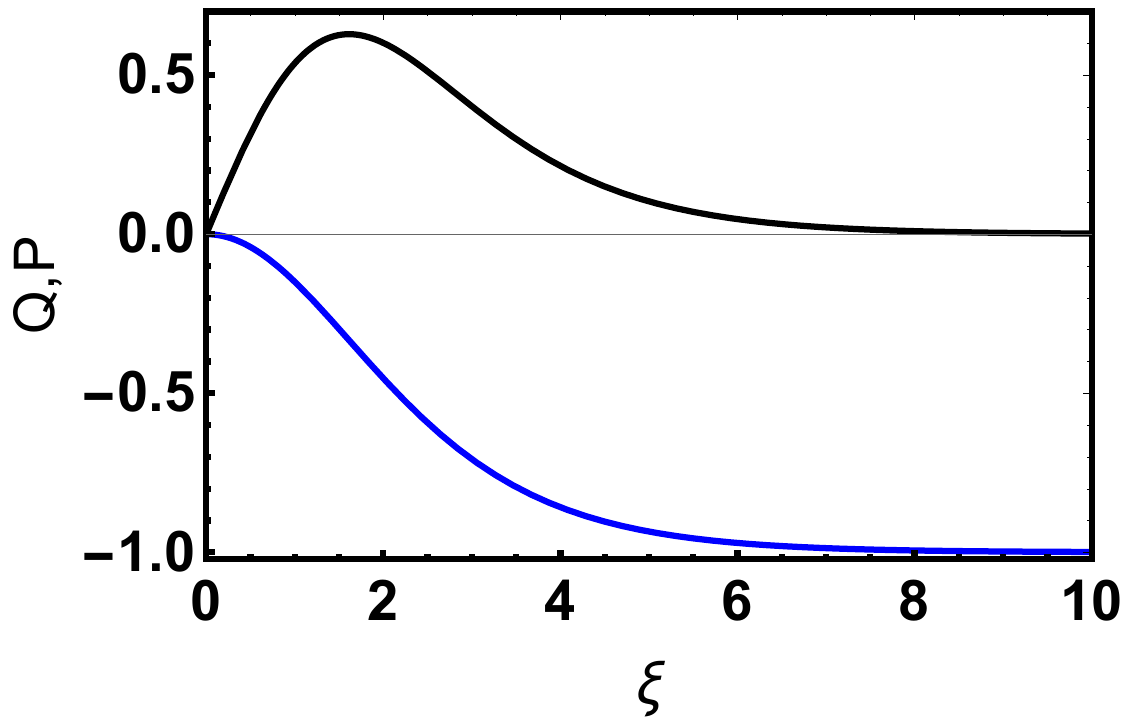}
\caption{The optimal solution conditioned on $c=0$. Shown are $Q(\xi)$ (black line) and $P(\xi)$ (blue line), as described by the exact Eqs.~(\ref{Qsimple}) and (\ref{Psimple}).}
\label{c=0_1}
\end{figure}

\begin{figure}
\includegraphics[width=0.4\textwidth,clip=]{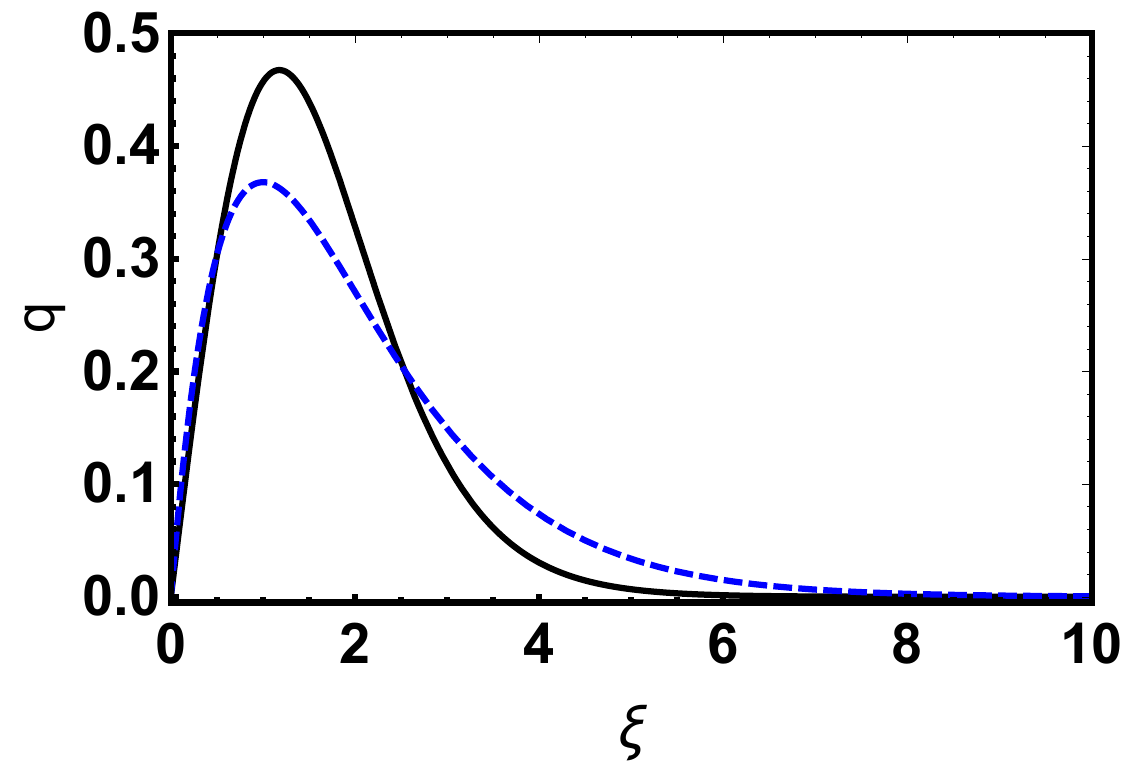}
\caption{Solid line: the optimal density profile $q(\xi)$, conditioned on $c=0$, as described by the exact Eq.~(\ref{qsimple}). Dashed line: the deterministic density profile $U(\xi) =\xi e^{-\xi}$, where $\xi =x-2t$.}
\label{c=0_2}
\end{figure}

\subsubsection{Numerical TFSs and critical point}
\label{numerics}

We solved Eqs.~(\ref{1B}) and~(\ref{2B}) numerically by the shooting method, that is by reducing the boundary value problem to an initial value problem, see \textit{e.g.} Ref. \cite{Stoer1980}.
Since, for given $P(\xi)$, Eq.~(\ref{1B}) for $Q(\xi)$ is linear and homogeneous, one can find the solution for an arbitrary positive value of $Q'(\xi=0)$, for example, for $Q'(\xi=0)=1$ and ultimately normalize the solution by using Eq.~(\ref{conservQ}). The rest of the initial conditions at $\xi=0$ are given by Eqs.~(\ref{BC0}) and (\ref{Pder}). The shooting parameter, at given $c$, is the Lagrange multiplier $\lambda$. It is determined either from the condition that, at large $\xi$, $P(\xi)$ approaches $-1$, or from the condition that $Q(\xi)$ obeys the asymptotic behavior (\ref{Qbehavior}). The former condition can be used only at $c>0$, because at $c<0$ $P(\xi)$ approaches $-1$ at any $\lambda$.
As an example, Fig. \ref{c=0_3} shows a comparison of the numerical solution for $Q(\xi)$ for $c=0$ with the exact analytical solution (\ref{Qsimple}). As one can see, the two graphs are indistinguishable.

\begin{figure}
\includegraphics[width=0.4\textwidth,clip=]{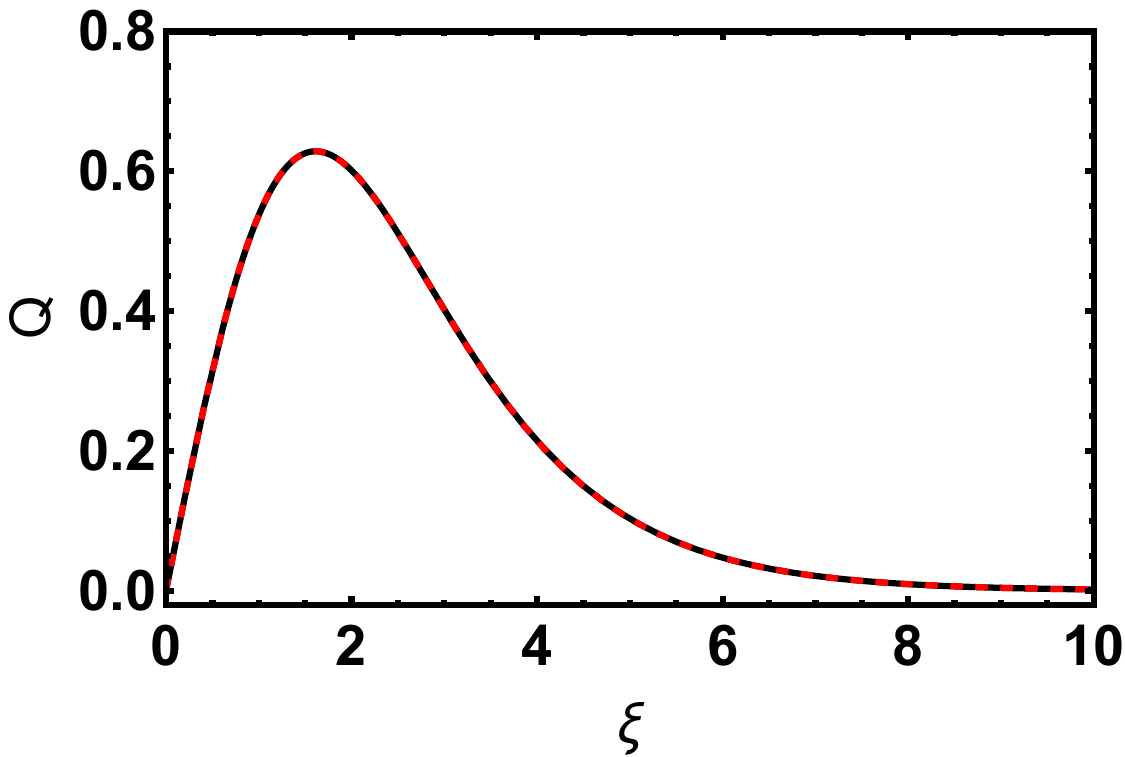}
\caption{Solid line: the numerical solution for $Q(\xi)$ for $c=0$. Dashed line: the exact solution (\ref{Qsimple}). The two curves are indistinguishable.}
\label{c=0_3}
\end{figure}

The blue line on Fig.~\ref{snum} shows the resulting rate function $s=\lambda$ as a function of c. One can notice that as $c<0$ decreases, $s(c)$ rapidly approaches the simple upper bound (\ref{sbound}). Qualitatively, this makes sense: the larger the displacement of the ensemble of particles ``in the wrong direction", the stronger the suppression of the branching process required.
What is remarkable, however, is that $s(c)$ reaches the bound at a finite value of $c=c_* = -3.38\dots$ which plays the role of a critical point \cite{accuratec*}.  At $c>c_*$ and close to $c_*$, the difference $\Delta=s_{\text{b}}(c)-s(c)$ [where $s_{\text{b}}(c)$ is described by Eq.~(\ref{sbound})]  is well fitted by a quadratic function:
\begin{equation}\label{subcritclose}
\Delta(c) \simeq  b (c-c_*)^2\,,
\end{equation}
where $b=0.047\dots$, see Fig.~\ref{Deltavsc}.  Importantly, there are no simple TFSs of the type (\ref{TFSsimple}) beyond the critical point, that is at $c<c_*$.

At the critical point $c=c_*$ the square roots in Eqs.~(\ref{Pdecay}) and (\ref{Qbehavior}) go away, and one obtains a limiting TFS with the asymptotics
\begin{eqnarray}
    1+P(\xi\to \infty) &\sim& \exp \left(\frac{c_*}{2}\xi\right)\,,\label{Plimitingdecay}\\
    Q(\xi\to \infty) &\sim& \exp\left(-\frac{c_*}{2}\xi\right) \label{Qlimitingbehavior}
\end{eqnarray}
(to remind the reader, $c_*<0$).  The resulting optimal particle density
$q(\xi)=Q(\xi) [1+P(\xi)]$ approaches a constant value at $\xi \to \infty$ which, in view of the normalization condition (\ref{conservQ}), must be set zero. Such an extreme delocalization of the density of this degenerate critical solution is the result of us taking the limit of $T\to \infty$.  It signals that, beyond the critical point, that is at $c<c_*$, the character of the true optimal solution must change. This is indeed what is observe in the supercritical solution that we present
in the next subsection.

\begin{figure}
\includegraphics[width=0.30\textwidth,clip=]{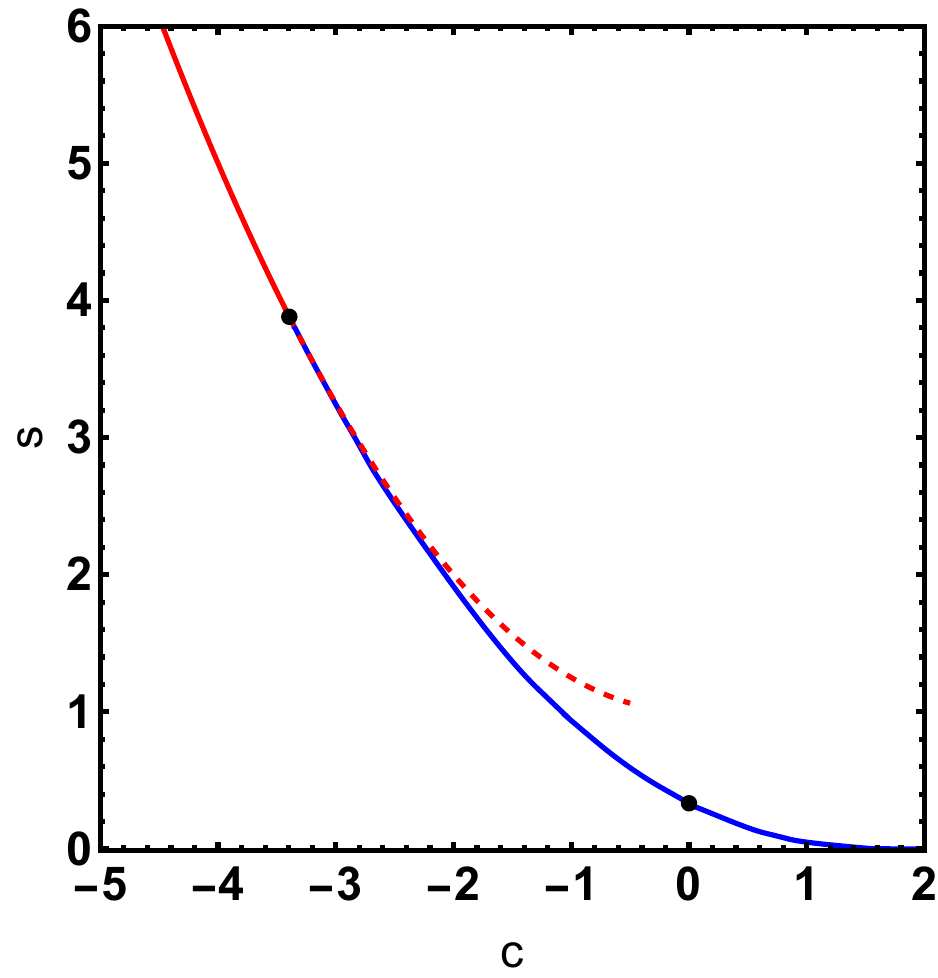}
\caption{The rate function per particle $s(c)$. In the subcritical region $c>c_* = -3.38\dots$,  $s(c)$ is determined numerically and shown by the blue solid line. The upper bound (\ref{sbound}) is shown by the solid and dashed red lines. The solid red line pertains to the supercritical region $c<c_*$ where the true $s(c)$, see Eq.~(\ref{snonstat}) below, coincides with the upper bound (\ref{sbound}).  The dashed red line pertains to the subcritical region. The two fat points show the exact result $s(0)=1/3$ and the numerical result $s(c_*) = 3.88\dots$.}
\label{snum}
\end{figure}

\begin{figure}
\includegraphics[width=0.32\textwidth,clip=]{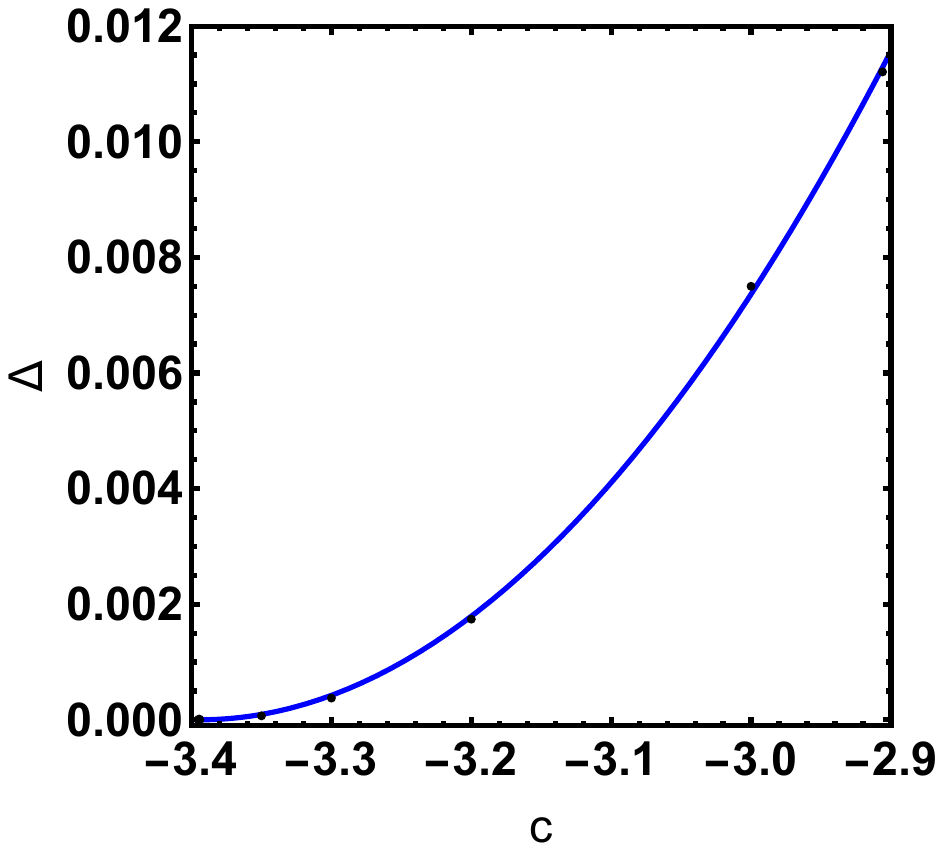}
\caption{$\Delta=c^2/4+1-s(c)$ vs. $c$ in the subcritical region $c>c_*=-3.38\dots$. The symbols show numerical results. The line is the quadratic fit (\ref{subcritclose}).}
\label{Deltavsc}
\end{figure}

\subsection{Asymptotic non-stationary solution for $c<c_*$}
\label{as}

At $c<c_*=-3.38\dots$ we must abandon the TFS ansatz~(\ref{TFSsimple}) and deal with the full time-dependent MFT problem, defined by Eqs.
(\ref{q1}), (\ref{p1}), (\ref{conservgeneral})-(\ref{probanonstat}) and (\ref{actionnonstat}). In particular, the simple relation (\ref{seqlambda}) does not hold here.

A guideline for establishing the character of solution comes from the fact that, for
the simple bound (\ref{sbound}) to apply, the branching of the particles must be strongly suppressed. This requires that, in the region where most of the particles are present, the terms $e^p$ in Eqs.~(\ref{q1}) and~(\ref{p1}) be negligible compared with the other terms. On the other hand, according to Eq.~(\ref{qat0}),  both $q$ and $p$ must vanish at the absorbing wall $x=X(t)$. As a result, there is some branching near the absorbing wall. As we will see, sufficiently close to the wall one has $p>0$, and the branching is even \emph{enhanced} compared to its mean-field behavior.

These arguments suggest that the absorbing wall moves faster than the particles' center of mass. This important feature is indeed present in the nonstationary MFT solution that we will now expose. In this long-time solution, $t\gg 1$, there is an extended \emph{bulk} region where $-p\gg 1$ and so the term $e^p$ is negligible, and a \emph{boundary layer} at the wall. This problem can be approximately solved via a matched asymptotic expansion because there is also a \emph{joint} region, where the bulk region and the boundary layer overlap. The bulk region obeys the condition $x-X(t)\gg 1$. The boundary layer is described by the condition $0<x-X(t)\ll\sqrt{t}$. In the joint region, where we can match the two asymptotics, one has  $1 \ll x-X(t)\ll\sqrt{t}$.

In the bulk region, where we can neglect the term $e^p$, Eq.~(\ref{p1}) simplifies to
\begin{equation}\label{N400}
\partial_tp=1-\left(\partial_xp\right)^2-\lambda(t)\,.
\end{equation}
We assume that the absorbing wall moves with a constant velocity, $X(t)=ct-at$, where $c$ is the constant velocity of the particles' center of mass as before, and the velocity difference $a=a(c)>0$ is \textit{a priori} unknown. With this assumption,  Eq.~(\ref{N400}) has the following simple solution:
\begin{equation}
  p^{\text{bulk}}(x,t) = \frac{c}{2}\left(x-ct+at\right)\,, \label{N410}
\end{equation}
whereas
\begin{equation}
\lambda = \frac{c^2}{4}+1-\frac{ca}{2} \label{N420}
\end{equation}
is independent of time. Now, plugging Eq.~(\ref{N410}) into Eq.~(\ref{q1}) and neglecting the term $e^p\ll1$, we obtain a simple linear equation for the density,
\begin{equation}\label{qlineq}
\partial_tq=\partial_x^2 q-c\partial_xq\,,
\end{equation}
which describes simple diffusion with a superimposed constant drift. The long-time solution of this equation is
\begin{equation}\label{N430}
q(x,t)=\frac{1}{\sqrt{4\pi t}}\exp\left[-\frac{(x-ct)^2}{4t}\right]\,.
\end{equation}
At this stage the positive constant $a$, entering the bulk solution (\ref{N410}) and (\ref{N430}) remains yet unknown.

Before proceeding to the boundary layer solution, let us determine the asymptotic behavior of the bulk density, as described by Eq.~(\ref{N430}), in the joint region. To this end let us transform Eq.~(\ref{N430}) to the frame, moving with the wall $x=X(t)$. That is, let us go over from $x$ to $\chi=x-X(t) = x-ct+at$. We obtain
\begin{eqnarray}
 q(\chi,t)&=& \frac{1}{\sqrt{4\pi t}}\exp\left[-\frac{(\chi-at)^2}{4t}\right] \nonumber\\
   &\equiv&  \frac{1}{\sqrt{4\pi t}}\exp\left(-\frac{\chi^2}{4t}+\frac{a\chi}{2}-\frac{a^2t}{4}\right)\,.\label{N435}
\end{eqnarray}
In the joint region, $\chi\ll \sqrt{t}$, the first term inside the exponent in the second line of Eq.~(\ref{N435}) can be neglected, and we obtain a separable asymptotic solution for the density in the joint region in the moving frame:
\begin{equation}
 q^{\text{joint}}(\chi,t)\simeq \frac{1}{\sqrt{4\pi t}}\exp\left(\frac{a\chi}{2}\right) \,\exp\left(-\frac{a^2t}{4}\right)\,.\label{jointq}
\end{equation}

Now we consider the boundary layer. Here $p(x,t)$ is described by a solution of Eq.~(\ref{p1}) which is stationary in the frame moving with the wall $X(t)$. This solution obeys the equation
\begin{equation}\label{N440}
e^{\tilde{p}}+\tilde{p}^{\prime\prime}-(c-a)\tilde{p}^\prime+\left(\tilde{p}^\prime\right)^2=
-\frac{c^2}{4}+\frac{ca}{2}\,.
\end{equation}
Here $p=\tilde{p}(\chi)$, and the primes denote the derivatives of functions of a single argument with respect to the argument.
We now turn to Eq.~(\ref{q1}) and transform from $x$ to $\chi$. We obtain the following equation for $q(\chi,t)$:
\begin{equation}\label{qbl}
\partial_t q -(c-a) \partial_{\chi} q = q e^{\tilde{p}}+\partial_{\chi}\left(\partial_{\chi}q-
2 q \tilde{p}'\right)\,.
\end{equation}
This partial differential equation is linear, and its coefficients are independent of time.
Only one of its separable solutions can be matched with the asymptotic (\ref{jointq}) in the joint region. This solution has the form
\begin{equation}\label{N460}
q(\chi,t)=R(\chi)\exp\left(-\frac{a^2 t}{4}\right)\,,
\end{equation}
where the function $R(\chi)$ obeys the equation
\begin{equation}\label{N450}
-\frac{a^2}{4}R-(c-a)R^\prime=e^{\tilde{p}}R+\left(R^\prime-2R\tilde{p}^\prime\right)^\prime\,.
\end{equation}

In order to complete the boundary layer solution, we need to solve the two ODEs~(\ref{N440}) and (\ref{N450}) for $\tilde{p}(\chi)$ and $R(\chi)$, respectively.  What are the boundary conditions for these two equations? For Eq.~(\ref{N440}) they are
\begin{equation}\label{N470}
\tilde{p}(0)=0\,,\quad
\tilde{p}(\chi\to\infty)\to\frac{c}{2}\chi\,.
\end{equation}
The condition at $\chi\to\infty$ is required for the matching, in the joint region, of the boundary layer solution for the conjugate momentum with the simple bulk solution $p(\chi) = (c/2)\chi$, see Eq.~(\ref{N410}).
In its turn, Eq.~(\ref{N450}) should be solved with the following boundary conditions:
\begin{equation}\label{N480}
R(0)=0\,,\quad
R(\chi\to\infty)\propto\exp\left(\frac{a\chi}{2}\right)\,.
\end{equation}
Again, the condition at $\chi\to\infty$ is  required in order to match the boundary layer solution -- this time for the density -- with the bulk solution in the joint region, see Eq.~(\ref{jointq}). Let us look at the behavior $R(\xi)$ at $\chi \to \infty$ more closely. As $\chi \to \infty$, the linear and homogeneous ODE~(\ref{N450}) simplifies to
\begin{equation}\label{Rinfty}
R''(\chi) -a R'(\chi) + \frac{a^2}{4} R(\chi)=0\,.
\end{equation}
The general solution of Eq.~(\ref{Rinfty}) is a linear combination of the two independent solutions $R_1(\chi) = \exp(a \chi/2)$ and $R_2(\chi) = \chi\,\exp(a \chi/2)$. According to the second boundary condition in Eq.~(\ref{N480}), the solution $R_2(\chi)$ must be suppressed which can be achieved by a proper choice of $a$ at given $c$. It is this condition that determines the velocity difference $a=a(c)>0$ between the TFS of $p$ (with the constant velocity $c-a$) and the center of mass of the particles.  The latter is travelling with the constant velocity $c$, in spite of the non-stationary character of the solution for optimal density.

We solved Eqs.~(\ref{N440}) and (\ref{N450}) for $\tilde{p}(\chi)$ and $R(\chi)$ numerically, by a shooting method, for $\chi<\chi_1$, where $\chi_1\gg1$ is a sufficiently large positive number.  The boundary conditions at $\chi=\chi_1$ are determined by the $\chi \to \infty$ asymptotic behaviors of $\tilde{p}(\chi)$ and $R(\chi)$ as described by Eqs.~(\ref{N470}) and~(\ref{N480}):
\begin{equation}\label{N500}
\tilde{p}(\chi_1)=\frac{c}{2}\chi_1\,,
\quad
\tilde{p}^\prime(\chi_1)=\frac{c}{2}\, ,
\end{equation}
\begin{equation}\label{N510}
R(\chi_1)=\alpha\,,
\quad
R^\prime(\chi_1)=\frac{a}{2}\alpha\, ,
\end{equation}
where the constant $\alpha$ in the r.h.s. of Eq.~(\ref{N510}) can be arbitrarily chosen by virtue of the linearity of Eq.~(\ref{N450}).
At specified $c<c_0$ the effective eigenvalue $a$ is determined by the condition that $\tilde{p}(\chi)$ and $R(\chi)$ vanish at the same point $\chi=\chi_0=O(1)$ which is \textit{a priori} unknown. The parameter $a$ serves as the shooting parameter. Once the solution is computed, it is shifted along the $\chi$-axis by $\chi_0$ so that the boundary conditions at $\chi=0$, see  Eqs.~(\ref{N470}) and~(\ref{N480}), are obeyed.

Figure \ref{BLsol} depicts a typical example of such a numerical solution.  As one can see,  $\tilde{p}>0$ near the wall, implying  an enhanced branching rate there. We observed this feature for all other values of $c<c_*$ that we worked with.

Figure \ref{avsc} shows the dependence of the velocity difference $a$ upon $c$ that we determined from these numerical solutions. $a$ vanishes at the critical point $c=c_*$ as to be expected. Sufficiently close to $c_*$ the function $a(c)$ appears to behave linearly with $c_*-c$.

\begin{figure}
\includegraphics[width=0.30\textwidth,clip=]{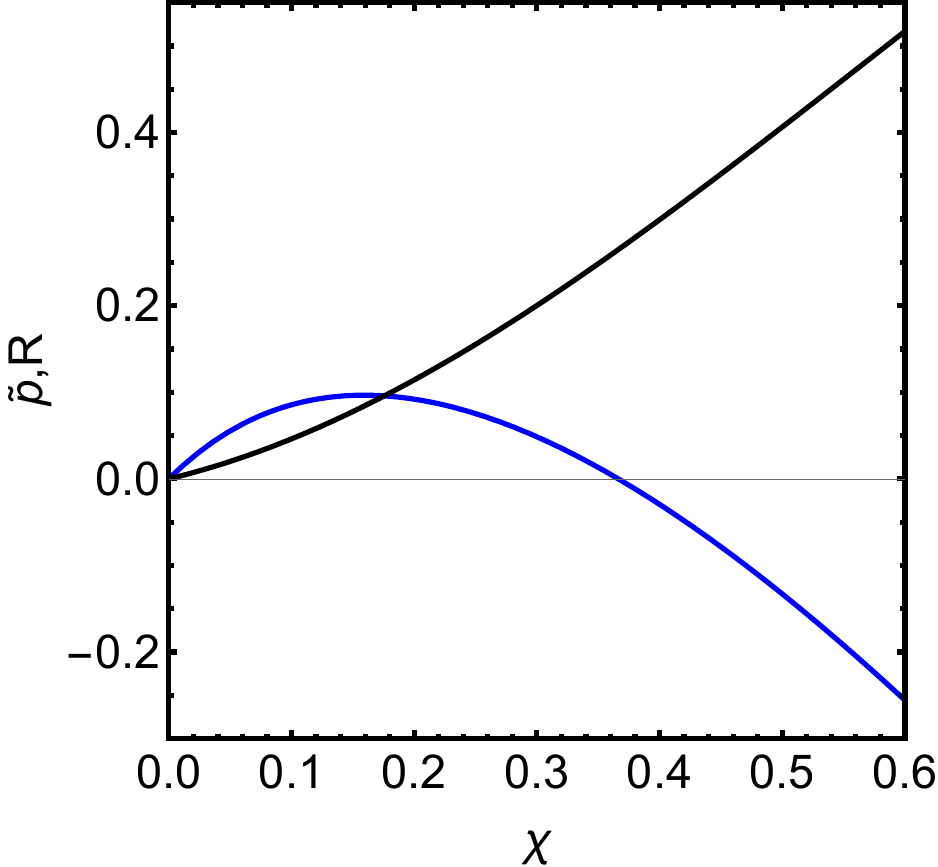}
\caption{Numerical solution in the boundary layer in the supercritical region $c<c_*=-3.38\dots$. Shown are $\tilde{p}(\chi)$ (blue) and $R(\chi)$ (black) for $c=-4$. In this case $a\simeq 0.347$.}
\label{BLsol}
\end{figure}

\begin{figure}
\includegraphics[width=0.30\textwidth,clip=]{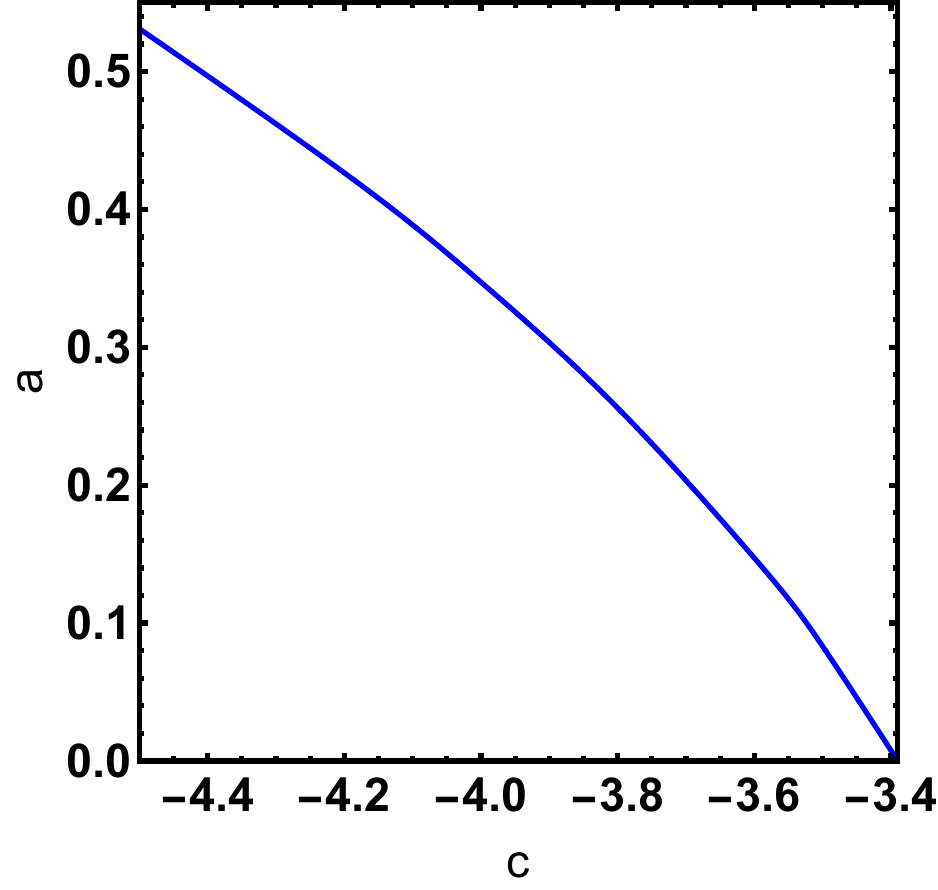}
\caption{The difference $a$  between the velocities of $p$ and of the center of mass of the system, is plotted as a function of $c$.}
\label{avsc}
\end{figure}

Now we turn to Eq.~(\ref{actionnonstat}) to calculate the rate function in the supercritical regime.
As $T\to \infty$, the dominating contribution to the action comes from the bulk region. Going over to the moving frame and neglecting the terms including $e^p$, we obtain in the leading order
\begin{equation}\label{actionbulk1}
 S(c<c_*,T)\simeq \int_0^T dt \int_{0}^{\infty}d\chi \,q(\chi,t) \{1+[\tilde{p}'(\chi)]^2 \}\,,
\end{equation}
As $\tilde{p}(\chi)\simeq c \chi/2$, this yields the rate function $s(c)$  at $N\to \infty$ and $T\to \infty$, which enters Eq.~(\ref{LDbehavior}):
\begin{equation}
\label{actionbulk2}
  s(c<c_*) = \frac{S(c,T)}{T}\Biggr|_{T\to\infty}= \left(\frac{c^2}{4}+1\right) \int_{0}^{\infty}d\chi \,q(\chi,t) \,.
\end{equation}
Using the normalization condition (\ref{conservgeneral}), we finally arrive at
\begin{equation}\label{snonstat}
s(c<c_*)=\frac{c^2}{4}+1\,,
\end{equation}
which perfectly coincides with the simple bound (\ref{sbound}). Note, that the knowledge of $a(c)$, which determines both the velocity difference between $p$ and $q$, and the Lagrange multiplier $\lambda(c)$ [see Eq.~(\ref{N420})], turns out to be unnecessary for the calculation of $s(c)$. The supercritical branch, $c<c_*$, of the rate function $s(c)$, is shown by the red solid line in Fig.~\ref{snum}.

\subsection{Dynamical phase transition}
We can summarize our results for the rate function $s(c)$ as follows:
\begin{equation}
\label{summarize}
s(c) =
\begin{cases}
\frac{c^2}{4}+1-b(c-c_*)^2+\dots,   & c>c_*\,,\\
\frac{c^2}{4}+1,  & c<c_*\,,
\end{cases}
\end{equation}
where $\dots$ denote higher-order terms in $c-c_*$. As one can see, at the critical point $c=c_*$, $s(c)$ is continuous together with its first derivative. The second derivative, however, experiences a jump. It is natural to classify this nonanalyticity of the rate function as a dynamical phase transition of second order.
The mechanism of this transition appears to be unrelated to any symmetry breaking. Rather, as we have seen, one type of solution for the optimal path -- a simple TFS -- disappears at the critical point, and another one -- the non-stationary solution -- appears.

\section{Discussion}
\label{discussion}

It was argued in Ref. \cite{MSF} that
close to $c=2$ the rate function $s(c)$ should be universal up to a model-dependent numerical constant for any microscopic reaction model where the binary branching $A\to 2A$ is the only first-order birth process. The $N$-BBM model belongs to this class, and our Eq.~(\ref{sc2}) supports the universality argument of Ref. \cite{MSF}.

Outside of the universal region $2-c\ll 2$, the optimal paths of the system and the ensuing rate function are model-dependent. We have shown that at $c_*<c<2$
the optimal path is described by the TFS ansatz~(\ref{TFSsimple}). We determined this TFS analytically for $c=0$ and numerically for other values of $c>c_*$.
The TFS ceases to exist at the critical point $c=c_*$, where it gives way a different solution. In this solution
the optimal conjugate momentum field $p$ still propagates as a TFS. However, the $p$-field, as well as a very small fraction of the particles,  travel faster than the center of mass of the system. Although center of mass does  move with a constant speed $c$, the density field itself is non-stationary and does not conform to any TFS. The presence of two different types of optimal solutions causes a second-order dynamical phase transition of the rate function at the critical point $c=c_*=-3.38\dots$.

In contrast to the universal dynamical phase transition at $c=2$ \cite{PTclarger2}, the second-order transition at $c=c_*=-3.38\dots$ is model-dependent. For comparison, in the stochastic FKPP problem, where the particles undergo the reversible reactions $A\to 2A$ and $2A\to A$ and Brownian motion, there is also a dynamical phase transition at $c<0$: in that case at $c=-2$  \cite{MSF}. However, the character of the optimal paths in the supercritical region is different there, as well as the order of that transition, which is infinite. It would be very interesting to look for additional examples of phase transitions at $c<0$ in models belonging to the FKPP universality class.

\section*{Acknowledgments}

B.M. is very grateful to Bernard Derrida for useful discussions and insightful comments and to Naftali R. Smith for advice. The research of B.M. was supported by the Israel Science Foundation (Grant No. 1499/20).


\begin{thebibliography}{99}
\bibitem{units} Suppose that the diffusion constant is $D$ and the branching rate is $k$. Then,
rescaling time by $1/k$ and the distance by $(D/k)^{1/2}$, one brings the model to a dimensionless form, where the diffusion constant and branching rate are equal to $1$.
\bibitem{othernames} The MFT is a variant of the optimal fluctuation method, also known
under the names of the weak-noise theory, dissipative WKB approximation,  and instanton method.
\bibitem{aboutB} Here the front position $u(\xi)$ is chosen uniquely so as to eliminate the subleading term of the asymptotic,  which behaves as $e^{-\xi}$. The coefficient $B\simeq 0.142$ is determined by solving the FKPP equation~(\ref{FKPP1}) numerically for $c=2$.
\end{thebibliography}
\end{document}